\documentclass[12pt]{article}
\usepackage{amsmath, amsthm,comment} 
\usepackage{amsfonts}
\usepackage{amssymb,graphics,psfrag}
\usepackage{array,epsfig,stmaryrd,graphicx}

\newcommand{\beq}{\begin{equation}}
\newcommand{\eeq}{\end{equation}}
\newcommand{\beqa}{\begin{eqnarray}}
\newcommand{\eeqa}{\end{eqnarray}}
\newcommand{\beqar}{\begin{eqnarray*}}
\newcommand{\eeqar}{\end{eqnarray*}}

\begin{document}
\baselineskip 18pt%
\begin{titlepage}
\vspace*{1mm}%
\hfill
\vbox{

    \halign{#\hfil         \cr
         CERN-PH-TH/2011-167\cr
           } 
      }  
\vspace*{10mm}
\vspace*{12mm}%

\center{ {\bf \Large  Critical Collapse in the Axion-Dilaton System
in Diverse Dimensions
 
}}\vspace*{3mm} \centerline{{\Large {\bf  }}}
\vspace*{5mm}
\begin{center}
{Luis \'Alvarez-Gaum\'e  and  Ehsan Hatefi}$\footnote{ 
On leave  from Department of Physics, Ferdowsi University of Mashhad, 
P.O. Box 1436, Mashhad, Iran.}$

\vspace*{0.8cm}{ {
Theory Group, Physics Department, CERN, CH-1211, Geneva 23, Switzerland}}
\vspace*{1.5cm}
\end{center}
\begin{center}{\bf Abstract}\end{center}
\begin{quote}
We study the gravitational collapse of the axion-dilaton system suggested by type IIB string
theory in dimensions ranging from four to ten.  We extend previous analysis concerning the role
played by the global $SL(2,R)$ symmetry and we evaluate the Choptuik exponents in the elliptic
case.
\end{quote}
\end{titlepage}%
\section{Introduction}

It is nearly twenty years since M. Choptuik discovered critical phenomena in gravitational collapse,
and revolutionised the field of numerical relativity \cite{Chop, gundlach}.  In this letter we revisit
the self-similar, spherical gravitational collapse studied in the past in four-dimensions in 
\cite{Eardley:1995ns, Hamade:1995ce}, and extend their analysis to higher dimensions and to other
implementations of self-similarity.  Our work is motivated by the AdS/CFT correspondence
\cite{maldacena, wittenone, klebanov, wittentwo}.  We are still far from a holographic description
of black hole formation in type IIB string theory.  Some exploratory work was done in \cite{scalingqcd,
AlvarezGaume:2008qs}.  In the context of type IIB (for reviews of string theory and duality see
\cite{greenschwarzwitten, polchinski}) and AdS/CFT one would like to consider collapse on
spaces which approach asymptotically $AdS_5\times S^5$.  With the bosonic fields in
the theory, a natural system to consider involves the axion-dilaton and the self-dual
$5$-form field.  One can show that in five dimensions the simplest dynamical set to study
involves just the Einstein-axion-dilaton system with a cosmological constant.  This poses
an apparent problem in considering self-similar collapse, because Einstein spaces do not admit homothetic
vector fields.  However, in the context of critical gravitational collapse we are considering
the collapse of matter to form small mass black holes and we only need to consider a small
space-time region close to where the singularity forms.  This should be independent
of the asymptotic structure of the space-time where the collapse takes place.  There is
numerical evidence in asymptotically AdS space-times showing that this is the case \cite{adscollapse}.
Hence we will eliminate the cosmological constant and analyse self-similar critical collapse
in dimensions $4-10$. We are interested in the critical solution and the corresponding 
Choptuik exponent. Our analysis follows closely \cite{Eardley:1995ns, Hamade:1995ce},
and the physical and geometrical interpretation is essentially the same, so we will not
review their work in this short letter.  

In section two we write the action and the equations of motion of the axion-dilaton system,
and the conditions that follow from the requirement of continuous self-similarity.  We found
three possible conditions and associated ans\"atze for the matter fields.  We called them
the elliptic, hyperbolic and parabolic cases associated to the three different clases of
$SL(2,R)$ transformations that can be used to compensate for a scaling transformation
in space-time.  The analysis in \cite{Eardley:1995ns, Hamade:1995ce} corresponds to the 
four-dimensional elliptic case, but their methods can easily be adapted to the
two other cases as well as to other dimensions.  We do not strive to obtain very precise
numerical results given that we have no practical applications for them.
We are interested in exploring the existence of critical solutions and the values
of their Choptuik exponents as a function of the space-time dimension.  For a massless
scalar field this has been done in \cite{dimdependence,Sorkin:2005vz}, the shape of the plot of
our exponents as a function of dimension is quite analogous to the results of Oren
and Sorkin in \cite{Sorkin:2005vz}, however more precise numerical work would be
necessary before a detailed comparison is possible.  More details as well as the
results for the hyperbolic and parabolic cases will appear in a subsequent publication
\cite{AlvarezGaume:2011ab}.

\section{The axion/dilaton system}

We will consider the $d$-dimensional spherical collapse of an axion-dilaton ($a,\Phi$)
system. Both field can be combined into a single complex scalar field
$ \tau \equiv a + i e^{- \phi}$ . 
The action for the model is:
\begin{equation}
S = {1\over 16 \pi G}\int d^d x \sqrt{-g} \left( R - {1 \over 2} { \partial_a \tau
\partial^a \bar{\tau} \over (\mathop{\rm Im}\tau)^2} \right) \; .
\label{eaction}
\end{equation}
where $R$ is the scalar curvature.  The equations of motion are:
\begin{eqnarray}
\label{eoms}
R_{ab} - {1 \over 4 (\mathop{\rm Im}\tau)^2} ( \partial_a \tau \partial_b
\bar{\tau} + \partial_a \bar{\tau} \partial_b \tau) & = & 0 
\end{eqnarray}
\begin{eqnarray}
\label{eoms14}
\nabla^a \nabla_a \tau + { i \nabla^a \tau \nabla_a \tau \over
\mathop{\rm Im}\tau} = 0 .
\end{eqnarray}
The theory is classically invariant under $SL(2,R)$ transformations:
\begin{equation}
\tau \rightarrow {a\tau+b \over c\tau+d} \; ,
\label{sltwo}
\end{equation}
where $(a,b,c,d) \in R$, $ad - bc = 1$ and $g_{ab}$ does not
transform.  This group is
supposed to be broken to its integer subgroup $SL(2,Z)$ as a consequence of
electric-magnetic duality in String Theory \cite{greenschwarzwitten, polchinski}.

Following \cite{Eardley:1995ns} the spherically symmetric metric can be written as:
\begin{equation}
	ds^2 = \left(1+u(t,r)\right)\left(- b(t,r)^2dt^2 + dr^2\right)
			+ r^2d\Omega^2_{d-2} \; .
\label{metric1}
\end{equation}
The time coordinate is chosen so that spherical collapse on the time axis
first occurs at $t=0$, hence the collapse takes place for $t<0$. We can still
implement time redefinitions for (\ref{metric1}), hence as in \cite{Eardley:1995ns}
we set $b(t,0)=1, t<0$ and regularity for $t<0$ implies that $u(t,0)=0, t<0$.  

Continuous self-similarity means the existence of a homothetic Killing vector
$\xi$ generating global scale transformations:
\begin{equation}
{\cal L}_{\xi} g_{ab} = 2 g_{ab} \; .
\label{metxi}
\end{equation}
In spherical coordinates $\xi=t\,\partial/\partial t+r\,\partial/\partial r$.
Defining the scale invariant variable $z=-r/t$ self-similarity of the metric
means that the unknown functions $u(t,r), b(t,r)$ are just functions of $z$.
The next question to address is the transformation of $\tau(t,r)$ under
scale transformations.  Since the action is $SL(2,R)$-invariant, we can
compensate a scale transformation of the coordinates $(t,r)$ by an $SL(2,R)$ transformation.
Thus if we change variables to $(t,z)$, we obtain a differential condition
for $\tau(t,z)$:
\begin{equation}
t\,{\partial\over \partial t}\,\tau(t,z)\,=\,\alpha_0\,+\,\alpha_1\,\tau\,+\,\alpha_2\,\tau^2
\end{equation}
with $\alpha_{0,1,2}$ real numbers. The quadratic polynomial on the right hand side
has two roots which can be two complex conjugate numbers, two distinct real numbers
or a double real root.  They correspond to compensating the scaling transformation
with respectively an elliptic, a hyperbolic, or a parabolic transformation in
$SL(2,R)$.  By straightforward manipulations and redefinitions the three cases
yield the following ans\"atze for $\tau(t,z)$.  In the elliptic case:
\begin{equation}
 \tau(t,r)	=  i { 1 - (-t)^{i \omega} f(z) \over 1 + (-t)^{i\omega} f(z)} ,
\label{tauansatz}
\end{equation}
under a scaling transformation $t\rightarrow \lambda\, t$, $\tau(t,r)$ changes
by a $SL(2,R)$ rotation.

In the hyperbolic case:
\begin{equation}
 \tau(t,r)	=   { 1 - (-t)^{ \omega} f(z) \over 1 + (-t)^{\omega} f(z)} ,
\label{tau2ansatz}
\end{equation}
and under a scaling transformation $t\rightarrow \lambda\, t$, $\tau(t,r)$ changes
by a $SL(2,R)$ boost.  Using a $SL(2,R)$-transformation we can transform 
$\tau(t,r)\rightarrow -(-t)^{\omega}\,f(z)$.
Finally, in the parabolic case a scaling transformation can be compensated by
a translation.  After some simple $SL(2,R)$ transformations of $\tau(t,r)$
the parabolic ansatz becomes:
\begin{equation}
\tau(t,r) = f(z)+\omega \log(-t)\; \;\;\;;\;\;\;w \in \mathbb{R}
\end{equation}

We have written $-t$ throughout the previous equations because the collapse will take place for
$t <0$, we can also extend the solutions for $t>0$, hence the correct ans\"atze for any value of
$t$ is to replace $-t\rightarrow |t|$.  In all these expressions, $f(z)$ is an arbitrary complex function
and $\omega$ a arbitrary real parameter to be fixed by requiring the critical solution to be
regular.

\section{Equations of Motion}

We collect in this section the equations of motion for the three ans\"atze, although
later we will only present results for the elliptic case.  Since we are implementing
spherical symmetry, the gravitational degrees of freedom do not propagate, there are
no gravitational waves.  Hence $u(z),b(z)$ should be expressed in terms of $f(z)$.
Common to all three cases, it is simple to show that 
\begin{equation}
u(z)\,=\,-{z\, b'(z)\over (d-3)\,b(z)}.
\end{equation}
This follows from the Einstein equation for the angular variables.
The other equations of motion for the self-similar solution involve $b(z), f(z)$.  In
fact the equation for $b$ is a first order linear inhomogeneous equation for $b(z)^2$
whose initial condition is determined by $b(0)=1$ together with the initial conditions 
for $f(z), f'(z)$, that are determined by requiring smoothness of the critical solution.
This determines not only the initial conditions but also the possible value of $w$.

The equations of motion are quite complicated, and their derivation is rather tedious
but straightforward.  The equations in the elliptic case are given by:
\begin{eqnarray}
0 & = & b' + { 2z(b^2 - z^2) \over (d-2)b (-1 + |f|^2)^2} f' \bar{f}' - {
2i \omega (b^2 - z^2) \over (d-2)b (-1 + |f|^2)^2} (f \bar{f}' - \bar{f} f')
- {2\omega^2 z |f|^2 \over (d-2)b (-1 + |f|^2)^2}, \nonumber\\
0 & = & f''
     - {2z (b^2 + z^2) \over (d-2)b^2 (-1 + |f|^2)^2} f'^2 \bar{f}'
     + {2 \over (1 - |f|^2)} \left(1
       - {i \omega (b^2 + z^2) \over (d-2) b^2 (1 - |f|^2)} \right) \bar{f} f'^2 \nonumber \\&&
     + {2i \omega (b^2 + 2 z^2) \over (d-2)b^2 (-1 + |f|^2)^2} f f'
\bar{f}' 
  + {2 \over z} \left(\frac{(z^2-\frac{(d-2)b^2}{2})}{(z^2-b^2)} + {i \omega z^2 (1 + |f|^2) \over (b^2 - z^2)
(1 - |f|^2)}\right.\nonumber \\&& 
+ \left.{2\omega^2 z^4 |f|^2 \over (d-2)b^2 (b^2 - z^2) (1 -
|f|^2)^2}\right) f'+ {2\omega^2 z \over (d-2)b^2 (-1 +|f|^2)^2} f^2
\bar{f}' + \nonumber \\&&
{2i \omega \over (b^2 - z^2)} \left(\frac{1}{2} - {i \omega (1 + |f|^2)
\over 2(1 - |f|^2)}\right.
- \left.{\omega^2 z^2 |f|^2 \over (d-2)b^2 (-1 + |f|^2)^2}
\right) f.
\label{1fzeom321}
\end{eqnarray}
It is useful to note that the equations are invariant under a global redefinition of
the phase of $f(z)$.  We can choose the phase of $f(z)$ to any convenient value at
a particular $z$ that we will choose to be the origin.

In the hyperbolic case, the equations of motion are quite similar, but now they
are invariant under a constant scaling $f\rightarrow \lambda f$, hence we can
choose $|f(z)|$ or its real or imaginary part as we wish at a particular value of $z$:
\begin{eqnarray}
0 & = & b' -{ 2z(b^2 - z^2) \over (d-2)b (f -\bar f)^2} f' \bar{f}' + {
2 \omega (b^2 - z^2) \over (d-2)b (f -\bar f)^2} (f \bar{f}'+ \bar{f} f')
+ {2\omega^2 z |f|^2 \over (d-2)b (f -\bar f)^2} \nonumber\\
0 & = & -f''
     - {2z (b^2 + z^2) \over (d-2)b^2 (f -\bar f)^2} f'^2 \bar{f}'
     + {2 \over (f -\bar f)} \left(\frac{1}{\bar f} 
       + { \omega (b^2 + z^2) \over (d-2) b^2 (f -\bar f)} \right) \bar{f} f'^2,\nonumber \\&&
     + {2 \omega (b^2 + 2 z^2) \over (d-2)b^2 (f -\bar f)^2} f f'
\bar{f}' 
  + {2 \over z} \left(-\frac{(z^2-\frac{(d-2)b^2}{2})}{(z^2-b^2)} + { \omega z^2 (f +\bar f) \over (b^2 - z^2)
(f -\bar f)}\right.\nonumber \\&& 
+ \left.{2\omega^2 z^4 |f|^2 \over (d-2)b^2 (b^2 - z^2) 
(f -\bar f)^2}\right) f'- {2\omega^2 z \over (d-2)b^2 (f -\bar f)^2} f^2
\bar{f}' + \nonumber \\&&
{2 \omega \over (b^2 - z^2)} \left(-\frac{1}{2} - { \omega (f +\bar f )
\over 2(f -\bar f)}\right.
- \left.{\omega^2 z^2 |f|^2 \over (d-2)b^2 (f -\bar f)^2}
\right) f.
\label{1fzeom321}
\end{eqnarray}

Finally in the parabolic ansatz the equations of motion are:
\begin{eqnarray}
0 & = & b'-{ 2z(b^2 - z^2) \over (d-2)b (f -\bar f)^2} f' \bar{f}' + {
2 \omega (b^2 - z^2) \over (d-2)b (f -\bar f)^2} ( \bar{f}'+ f')
+ {2\omega^2 z  \over (d-2)b (f -\bar f)^2}, \nonumber\\
0 & = & -f''
     - {2z (b^2 + z^2) \over (d-2)b^2 (f -\bar f)^2} f'^2 \bar{f}'
     + {2 \over (f -\bar f)} \left(1 
       + { \omega (b^2 + z^2) \over (d-2) b^2 (f -\bar f)} \right) f'^2 \nonumber \\&&
     + {2 \omega (b^2 + 2 z^2) \over (d-2)b^2 (f -\bar f)^2}  f'
\bar{f}' 
  + {2 \over z} \left(-\frac{(z^2-\frac{(d-2)b^2}{2})}{(z^2-b^2)} + {2\omega z^2  \over (b^2 - z^2)
(f -\bar f)}\right.\nonumber \\&& 
+ \left.{2\omega^2 z^4  \over (d-2)b^2 (b^2 - z^2) 
(f -\bar f)^2}\right) f'- {2\omega^2 z \over (d-2)b^2 (f -\bar f)^2} 
\bar{f}' + \nonumber \\&&
{2 \omega \over (b^2 - z^2)} \left(-\frac{1}{2} - { \omega
\over (f -\bar f)}\right.
- \left.{\omega^2 z^2  \over (d-2)b^2 (f -\bar f)^2}
\right).
\label{1fzeom321}
\end{eqnarray}
They are invariant under arbitrary shifts of $f(z)$ by a real number, and thus we can choose
its real part as we wish at a particularly convenient point.
\begin{table}[t]
\centering
\begin{tabular}{cccc}
\hline
\quad\quad\quad\quad\quad$\omega$\quad\quad\quad$z_{+}$\quad\quad$|f(0_{+})|$\quad\quad$|f(z_{+})|$\quad\quad
 \\
\hline
\quad1.176\quad\quad$2.609$\quad\quad.892\quad\quad$.364$\quad\quad
\\
\\
\quad1.297\quad\quad$2.674$\quad\quad.903\quad\quad$.392$\quad\quad
\\
\\
\quad1.469\quad\quad$2.771$\quad\quad.910\quad\quad$.397$\quad\quad
\\
\\
\quad1.610\quad\quad$2.822$\quad\quad.914\quad\quad$.404$\quad\quad
\\
\\
\quad1.721\quad\quad$2.871$\quad\quad.918\quad\quad$.413$\quad\quad
\\
\\
\quad1.791\quad\quad$2.929$\quad\quad.923\quad\quad$.427$\quad\quad
\\
\\
\quad1.852\quad\quad$2.998$\quad\quad.928\quad\quad$.439$\quad\quad
\\
\\
\hline
\end{tabular}
\caption{Parameters determining the critical solution for dimensions $4-10$, with dimension increasing
from top to bottom.}
\label{qmax}
\end{table}

\section{Properties of the critical solutions in the elliptic case and their Choptuik exponents}

We briefly present some of our results in the study of the critical solution in the elliptic case.
The analysis follows closely the arguments in \cite{Eardley:1995ns, Hamade:1995ce}.  In all three non-linear
systems (12-14) we have five singular points, $z=\pm 0$ represents the axis $r=0$ and regularity is imposed.
The point $z=\infty$ represents the surface $t=0$.  Away from $r=0$ there is nothing special in this
surface, hence regularity is also imposed.  The simplest way to see that there is no problem at $z=\infty$
is through a change of variables and a redefinition of the fields $f(z),b(z)$ \cite{Eardley:1995ns}.
The singularities $b(z_{\pm})=\pm z_{\pm}$ are those surfaces where the homothetic Killing vector becomes null.  They correspond to the backward (resp. forward) light cone of the spacetime origin.  For $b(z_+)=z_+$ the
solution should be smooth across this surface.  The forward cone $b(z_-)=-z_-$ of the singularity represents
the Cauchy horizon and we should not require more than continuity of $f,b$ across this surface.  We
require smoothness of the spacetime below the forward cone and then extend the space by continuity
inside it.  In the elliptic case it is convenient to write $f(z)=f_m(z) e^{i f_a(z)}$, requiring regularity
both at the origin and at $z_+$ leaves only four parameters to be determined: $|f(0)|,w,z_+, |f(z_+)|$.  To
find them we proceed by integrating from close to the origin towards $z_+$ and also from  a little
below $z_+$ towards the origin, and match the results at an intermediate point, say $z=1$.  By matching the
functions and the relevant first derivatives we can determine the unknown parameters in the critical 
solution.  We do this for every dimension between $d=4$ and $d=10$.  The results are shown in Table 1.
We did not optimise the numerical precision for these computations.  In determining the analogous values
of $z_-,f(z_-)$ more precision is required because the solution is quite flat inside the forward light
cone.  The results will appear in \cite{AlvarezGaume:2011ab}.

We compute the Choptuik exponents following the methods in \cite{koike, maison, gundlach, Hamade:1995ce}.
Given the critical solution, we can perturb it
to find the critical exponent $\gamma$ as
follows. Let $h$ be any function determining the critical solution: $b$ or $f$. 
Next consider small perturbations around the critical solution: 
\begin{eqnarray}
h(z,t) = h_{\rm ss}(z) + \epsilon |t|^{- \kappa} h_{\rm pert}(z) 
\label{decay}
\end{eqnarray}
where $h_{\rm ss}(z)$ is the critical solution, $\epsilon$ is a small
number, $\kappa$ is a constant, and $h_{\rm pert}(z)$ depends only on
$z$. Substituting  $h(z,t)$ in the full equations of motion of motion (2,3) and keeping
first order terms in $\epsilon$, gives a set of linear equations for 
the perturbation and an eigenvalue equation for
$\kappa$. This eigenvalue equation can in principle have a number of
possible solutions for $\kappa$. The solution with the largest value
of $Re (\kappa)$ will be responsible for the fastest growing perturbation in  (\ref{decay}),
and is called the ``most relevant mode.'' The critical exponent is
given by
\begin{eqnarray}
\gamma = \frac{1}{Re(\kappa)}
\end{eqnarray}
We require also that the perturbations are smooth at the singularities of 
the original equations.  This determines the $h_{\rm pert}(z)$ up to some rescalings
(which is fine given that the equations are linear) and also the most relevant
value for ${\rm Re}(\kappa)$.  In fact, we take $\kappa$ to be real.
The values we obtain for the Coptuik exponents in dimensions from $4-10$ appear
in Table 2.  The numerical accuracy decreases as the dimension increases.  The
third significant digit in Table 2 should not be trusted, specially in dimensions
$9$ and $10$.  More powerful numerical methods are necessary to get more accurate
values for the exponents.  
\begin{table}
\centering
\begin{tabular}{cc}
\hline
\quad\quad\quad${\rm dimension}$\quad\quad\quad$\gamma$ 
\\
\hline
\quad\quad\quad\quad$4$\quad\quad\quad\quad\quad$.261$
\\
\quad\quad\quad\quad$5$\quad\quad\quad\quad\quad$.278$
\\
\quad\quad\quad\quad$6$\quad\quad\quad\quad\quad$.287$
\\
\quad\quad\quad\quad$7$\quad\quad\quad\quad\quad$.311$
\\
\quad\quad\quad\quad$8$\quad\quad\quad\quad\quad$.352$
\\
\quad\quad\quad\quad$9$\quad\quad\quad\quad\quad$.373$
\\
\quad\quad\quad\quad$10$\quad\quad\quad\quad\quad$.347$
\\
\hline
\end{tabular}
\caption{Choptuik exponents in dimensions $4-10$ for the elliptic case.}
\label{qmax}
\end{table}
We have preliminary results in the hyperbolic and parabolic cases indicating
that there are also critical solutions in diverse dimensions with Choptuik
exponents different from those found for the elliptic case \cite{AlvarezGaume:2011ab}.

As mentioned in the introduction the $SL(2,R)$ symmetry of the classical type IIB
string theory is supposed to break to $SL(2,\mathbb{Z})$ once quantum effects are taken into
account.  This raises the very interesting possibility that the critical solution in
the quantum case will not be continuous self-similar but rather it will have discrete
self-similarity as for the massless scalar field first analyzed by Choptuik \cite{Chop}.
In other words, we want to know if there are elements $\Gamma\in SL(2,\mathbb{Z})$, such that:
\begin{equation}
\tau(e^{\Delta_\Gamma}\,t,e^{\Delta_\Gamma}\, r)\,=\,{a\tau+b\over c\tau+d};\qquad \Gamma\,=\,\left(
\begin{array}{cc}a & b \\
c & d \end{array} \right)\in SL(2,\mathbb{Z}),
\end{equation}
with $\Delta_{\Gamma}$ the corresponding echo parameter.  To settle this question one would
have to do the full numerical integration of Einstein's equation without assuming continuous
self-similarity. 
\section*{Acknowledgment}
E.H acknowledges the theory division of CERN for its hospitality, he also thanks E.Hirschmann for a private communication.  We also thankA. Sabio-Vera,  M.A. V\'azquez-Mozo and A. Tavanfar for discussions.



\begin{thebibliography}{2007}
\bibitem{Chop} M.W.~Choptuik, 1993,
 Universality and Scaling in Gravitational Collapse of a Massless Scalar Field\ 
{\it Phys.\ Rev. Lett.}\ {\bf70}, 9-12

\bibitem{gundlach}
For a review and references, see:
C. Gundlach, 2002, Critical phenomena in gravitational collapse, {\it Phys.\ Rept.\ } {\bf 376} 339-405 
(arXiv:gr-qc/0210101)
  
 \bibitem{Eardley:1995ns}
  D.~M.~Eardley, E.~W.~Hirschmann and J.~H.~Horne,1995\ 
S duality at the black hole threshold in gravitational collapse,
  {\it Phys.\ Rev.}\  D {\bf 52} (1995) 5397
  (arXiv:gr-qc/9505041)
  
\bibitem{Hamade:1995ce}
  R.~S.~Hamade, J.~H.~Horne and J.~M.~Stewart\  1995 
  Continuous Self-Similarity and $S$-Duality,
  {\it Class.\ Quant.\ Grav.\ } {\bf 13} (1996) 2241-2253
  [arXiv:gr-qc/9511024].
  
  \bibitem{maldacena}
Maldacena J M 1998 The large N limit of superconformal field theories and supergravity {\it Adv.\ Theor.\ Math.\
Phys.\ } {\bf 2} 231Ð52 (arXiv:hep-th/9711200)

\bibitem{wittenone}
Witten E 1998 Anti-de Sitter space and holography {\it Adv.\ Theor.\ Math.\ Phys.\ }{\bf 2} 253Ð91 (arXiv:hep-th/9802150)

\bibitem{klebanov}
Gubser S S, Klebanov I R and Polyakov A M 1998 Gauge theory correlators from non-critical string theory
{\it Phys.\ Lett.\ }{\bf B 428} 105Ð14 (arXiv:hep-th/9802109)

\bibitem{wittentwo}
Witten E 1998 Anti-de Sitter space, thermal phase transition, and confinement in gauge 
theories {\it Adv.\ Theor.\ Math.\ Phys.\ } {\bf 2} 505Ð32 (arXiv:hep-th/9803131)

\bibitem{scalingqcd}
L.~\'Alvarez-Gaum\'e, C.~G\'omez, and M.~A.~V\'azquez-Mozo,\ 2007 Scaling Phenomena in Gravity from QCD
{\it Phys.\ Lett. \ }{\bf B649} 478-482 (arXiv: hep-th/0611312)

\bibitem{AlvarezGaume:2008qs}
  L.~\'Alvarez-Gaum\'e, C.~G\'omez, A.~Sabio Vera, A.~Tavanfar and M.~A.~V\'azquez-Mozo,\  2008
  Critical gravitational collapse: towards a holographic understanding of the
  Regge region,
  {\it Nucl.\ Phys.\ } B {\bf 806}, 327-385 
  (arXiv:0804.1464 [hep-th])
  
\bibitem{greenschwarzwitten}
M.B. Green, J.H. Schwarz and E. Witten,\ 1987 {\it Superstring Theory} Vols I,II, Cambridge University Press, 

\bibitem{polchinski}
J. Polchinski,\ 1998 {\it String Theory}, Vols I,II, Cambridge University Press 

\bibitem{adscollapse}
M. Birukou, V. Husain, G. Kunstatter, E. Vaz, M. Olivier, \ 2002
Spherically symmetric scalar field collapse in any dimension,
{\it Phys. \ Rev.\ } {\bf D 65} 104036 (arXiv:gr-qc/0201026), and

V. Husain, G. Kunstatter, B. Preston, M. Birukou, 2003
Anti-de Sitter gravitational collapse,
{\it Class.\ Quant.\ Grav. \ } {\bf 20}:L23-L30

\bibitem{dimdependence}
J. Bland, B. Preston, M. Becker, G. Kunstatter,  V. Husain, \ 2005
Dimension dependence of the critical exponent in spherically symmetric gravitational collapse,
{\it Class.\ Quant.\ Grav.\ }{\bf 22}\, 5355-5364 

\bibitem{Sorkin:2005vz}
  E.~Sorkin and Y.~Oren, \ 2005
  On Choptuik's scaling in higher dimensions,
 {\it Phys.\ Rev.}\  D {\bf 71}, 124005 
  (arXiv:hep-th/0502034)

\bibitem{AlvarezGaume:2011ab}
  L.~\'Alvarez-Gaum\'e and E. Hatefi, in preparation.
  
\bibitem{koike}
 T.~Koike, T.~Hara and S.~Adachi,\ 1995\ 
 Critical behavior in gravitational collapse of radiation fluid: A
  Renormalization group (linear perturbation) analysis
  Phys.\ Rev.\ Lett.\  {\bf 74}, 5170 
  (arXiv:gr-qc/9503007).

\bibitem{maison}
  D.~Maison,\ 1996\ 
Nonuniversality of critical behavior in spherically symmetric gravitational
  collapse
  Phys.\ Lett.\  B {\bf 366}, 82 (1996)
  (arXiv:gr-qc/9504008).

  







  \end{thebibliography}
\end{document}